\documentclass[10pt,aps,prc,twocolumn,superscriptaddress,showpacs,floatfix,nofootinbib]{revtex4-1}

\usepackage{array}
\usepackage{amsmath}
\usepackage{mathptmx}
\usepackage{mathptm}
\usepackage{amssymb}
\usepackage{bm}
\usepackage{wasysym}
\usepackage{multirow}
\usepackage{float}
\usepackage[final]{graphics}
\usepackage{epstopdf}
\usepackage{placeins}
\usepackage{xcolor}
\usepackage{subfloat}
\usepackage{rotating}
\usepackage{booktabs}
\usepackage{tabularx}
\usepackage[caption=false]{subfig}
\usepackage[sort&compress]{natbib}
\usepackage{fontenc}
\usepackage{times}
\usepackage{epsfig}
\usepackage{verbatim}
\usepackage{threeparttable}
\usepackage{dcolumn}
\newcolumntype{.}{D{.}{.}{7}}
\usepackage[english]{babel}
\usepackage{chngcntr}
\usepackage{hycolor}
\makeatletter

\begin{document}

%%%%%%%%%%%%%%%%%%%%%%%%%%%%%%%%%%%%%%%%%%%%%%%%%%%%%%%%%%%%%%%%%%%%%%%%%%%%%%%%%%%%%%%%%%%%%%%%%%%%%%%%%
%% Remove these lines before the final submission to remove line numbers at the beginning of each line. %
%\setpagewiselinenumbers                                                                                %
%\modulolinenumbers[1]                                                                                  %
%\linenumbers                                                                                           %
%%%%%%%%%%%%%%%%%%%%%%%%%%%%%%%%%%%%%%%%%%%%%%%%%%%%%%%%%%%%%%%%%%%%%%%%%%%%%%%%%%%%%%%%%%%%%%%%%%%%%%%%%

%%%%%%%%%%%%%%%%%%%%%%%%%%%%%%
%% Author and Collaborators %%
%%%%%%%%%%%%%%%%%%%%%%%%%%%%%%

\title{Excited states of $^{39}$Ca and their significance in nova nucleosynthesis}
\author{K.~Setoodehnia}
 \email{ksetood@ncsu.edu}
 \affiliation{Department of Physics, North Carolina State University, Raleigh NC, 27695, USA\\}
 \affiliation{Triangle Universities Nuclear Laboratory, Duke University, Durham NC, 27710, USA\\}
\author{C.~Marshall}
 \affiliation{Department of Physics, North Carolina State University, Raleigh NC, 27695, USA\\}
 \affiliation{Triangle Universities Nuclear Laboratory, Duke University, Durham NC, 27710, USA\\}
\author{R.~Longland}
 \affiliation{Department of Physics, North Carolina State University, Raleigh NC, 27695, USA\\}
 \affiliation{Triangle Universities Nuclear Laboratory, Duke University, Durham NC, 27710, USA\\}
\author{J.~H.~Kelley}
 \affiliation{Department of Physics, North Carolina State University, Raleigh NC, 27695, USA\\}
 \affiliation{Triangle Universities Nuclear Laboratory, Duke University, Durham NC, 27710, USA\\}
\author{J.~Liang}
 \affiliation{Department of Physics \& Astronomy, McMaster University, Hamilton ON, L8S 4M1, Canada\\}
\author{F.~Portillo Chaves}
 \affiliation{Department of Physics, North Carolina State University, Raleigh NC, 27695, USA\\}
 \affiliation{Triangle Universities Nuclear Laboratory, Duke University, Durham NC, 27710, USA\\}

%%%%%%%%%%
%% Date %%
%%%%%%%%%%

\date{\today}

%%%%%%%%%%%%%%
%% Abstract %%
%%%%%%%%%%%%%%

\begin{abstract}
\textbf{Background:} Discrepancies exist between the observed abundances of argon and calcium in oxygen-neon nova ejecta and those predicted by nova models. An improved characterization of the $^{38}$K($p, \gamma$)$^{39}$Ca reaction rate over the nova temperature regime ($\sim$ 0.1 -- 0.4 GK), and thus the nuclear structure of $^{39}$Ca above the proton threshold (5770.92(63) keV), is necessary to resolve these contradictions. \textbf{Purpose:} The present study was performed to search for low-spin proton resonances in the $^{38}$K $+$ $p$ system, and to improve the uncertainties in energies of the known astrophysically significant proton resonances in $^{39}$Ca. \textbf{Method:} The level structure of $^{39}$Ca was investigated via high-resolution charged-particle spectroscopy with an Enge split-pole spectrograph using the $^{40}$Ca($^{3}$He, $\alpha$)$^{39}$Ca reaction. Differential cross sections were measured over 6 laboratory angles at 21 MeV. Distorted-wave Born approximation calculations were performed to constrain the spin-parity assignments of observed levels with special attention to those significant in determination of the $^{38}$K($p, \gamma$)$^{39}$Ca reaction rate over the nova temperature regime. \textbf{Results:} The resonance energies corresponding to two out of three astrophysically important states at 6154(5) and 6472.2(24) keV are measured with better precision than previous charged-particle spectroscopy measurements. A tentatively new state is discovered at 5908(3) keV. The spin-parity assignments of a few of the astrophysically important resonances are determined. \textbf{Conclusions:} The present $^{38}$K($p, \gamma$)$^{39}$Ca upper limit thermonuclear reaction rate at 0.1 -- 0.4 GK is higher than that determined in [Physical Review C 97 (2018) 025802] by at most a factor of 1.4 at 0.1 GK.
\end{abstract}

%%%%%%%%%%%%%%%%%
%% Pac Numbers %%
%%%%%%%%%%%%%%%%%

\pacs{26.30.Ca,25.40.Hs,27.30.+Z,29.30.Ep,29.40.Gx}
\maketitle

%%%%%%%%%%%%%%%
%% Main Text %%
%%%%%%%%%%%%%%%

\section{\label{Introduction}Introduction}

Classical novae occur in semi-detached~\cite{Renvoize:2002} close interacting binary systems consisting of a Carbon-Oxygen-rich (CO) or an Oxygen-Neon-rich (ONe) white dwarf and a main sequence star. When the white dwarf has accreted sufficient hydrogen-rich material ($\sim$10$^{-4}$ $M_{\astrosun}$ to 10$^{-6}$ $M_{\astrosun}$~\cite{Starrfield:2016}) from its companion star, a thermonuclear runaway occurs on the surface of the white dwarf at the base of the accreted envelope (see Ref.~\cite{Jose:2016} for details). Depending on the mass of the white dwarf, peak temperatures of 0.1 -- 0.4 GK are reached within only a few hundred seconds or less~\cite{Starrfield:2016}. The dominant nuclear reaction flow proceeds through explosive hydrogen burning via the $rp$-process~\cite{Bertulani:2016}, where a series of ($p, \gamma$) and ($p, \alpha$) reactions and $\beta^{+}$-decays occur on the proton-rich side of the valley of stability. An outburst follows giving birth to a classical nova.\par
Relative to solar abundances, the ejecta of classical novae show significant nuclear processing~\cite{Jose:1998}. Both theoretical estimations~\cite{Iliadis:2002,Denissenkov:2014,Starrfield:2016,Jose:2017,Figueira:2018} and the abundance patterns inferred from observations of nova ejecta~\cite{Andrea:1994,Pottasch:1959,Snijders:1987,Morisset:1996,Arkhipova:2000,Gehrz:2002,Evans:2003,Helton:2012} agree that the nuclear activity in classical novae generally stops around $A$ $\sim$ 40, i.e., calcium.\par
Observed elemental abundances for Ar and Ca in ONe novae~\cite{Andrea:1994} show an enhancement with respect to the solar abundances by up to an order of magnitude. Whereas, nova models~\cite{Starrfield:2009} predict such abundances to be generally close to the solar values. A sensitivity study performed by Iliadis {\em et al.}~\cite{Iliadis:2002} identified the $^{38}$K($p, \gamma$)$^{39}$Ca reaction as one of the few reactions affecting the simulated nova abundances for Ar and Ca.\par
The $^{38}$K($p, \gamma$)$^{39}$Ca reaction rate of Ref.~\cite{Iliadis:2002} was determined solely based on Hauser-Feshbach calculations, which introduced 4 orders of magnitude uncertainty in the reaction rate. Consequently, variations of factors of $\sim$25, 136 and 58 were found in the predicted $^{38}$Ar, $^{39}$K and $^{40}$Ca final abundances, respectively~\cite{Iliadis:2002}.\par
Over the temperatures characteristic of explosive hydrogen burning in novae, the $^{38}$K($p, \gamma$)$^{39}$Ca reaction rate is
dominated by contributions from three \textit{l} $=$ 0, $J^{\pi}$ $=$ 5/2$^{+}$ states in $^{39}$Ca. The excitation energies of these states were
reported~\cite{Singh:2006} to be 6157(10), 6286(10) and 6460(10) keV based on various previous measurements. They correspond to proton resonances in the $^{38}$K $+$ $p$ system ($Q$ $=$ 5770.92(63) keV~\cite{Wang:2017}) at $E_{r}$ $=$ 386(10), 515(10) and 689(10) keV, respectively. In order to reduce the uncertainty in the $^{38}$K($p, \gamma$)$^{39}$Ca reaction rate to help constrain the nova models and remove the discrepancy
between simulated and observed Ar and Ca abundances in nova ejecta, the $^{38}$K($p, \gamma$)$^{39}$Ca reaction was directly measured in inverse kinematics recently for the first time~\cite{Lotay:2016,Christian:2018}.\par
As a result, the 689(10)-keV resonance in $^{39}$Ca was observed at $E_{r}$ $=$
679$^{+2}_{-1}$(stat.) $\pm$1(sys.) keV, and its strength was measured to be $\omega\gamma$ $=$ 120$^{+50}_{-30}$(stat.)$^{+20}_{-60}$(sys.) meV. The other two resonances remained unobserved~\cite{Christian:2018}; however, upper limits on their strengths were determined. The $^{38}$K($p, \gamma$)$^{39}$Ca reaction rate was recalculated and its uncertainty was reduced to a factor of $\sim$40~\cite{Christian:2018}. This, in turn, has reduced the uncertainty in predicted abundances of $^{38}$Ar, $^{39}$K and $^{40}$Ca to a factor of $\leq$ 15~\cite{Christian:2018}.\par
Further high-resolution spectroscopic studies of $^{39}$Ca were

\begin{figure*}[ht]
\begin{center}
\includegraphics[width=\textwidth]{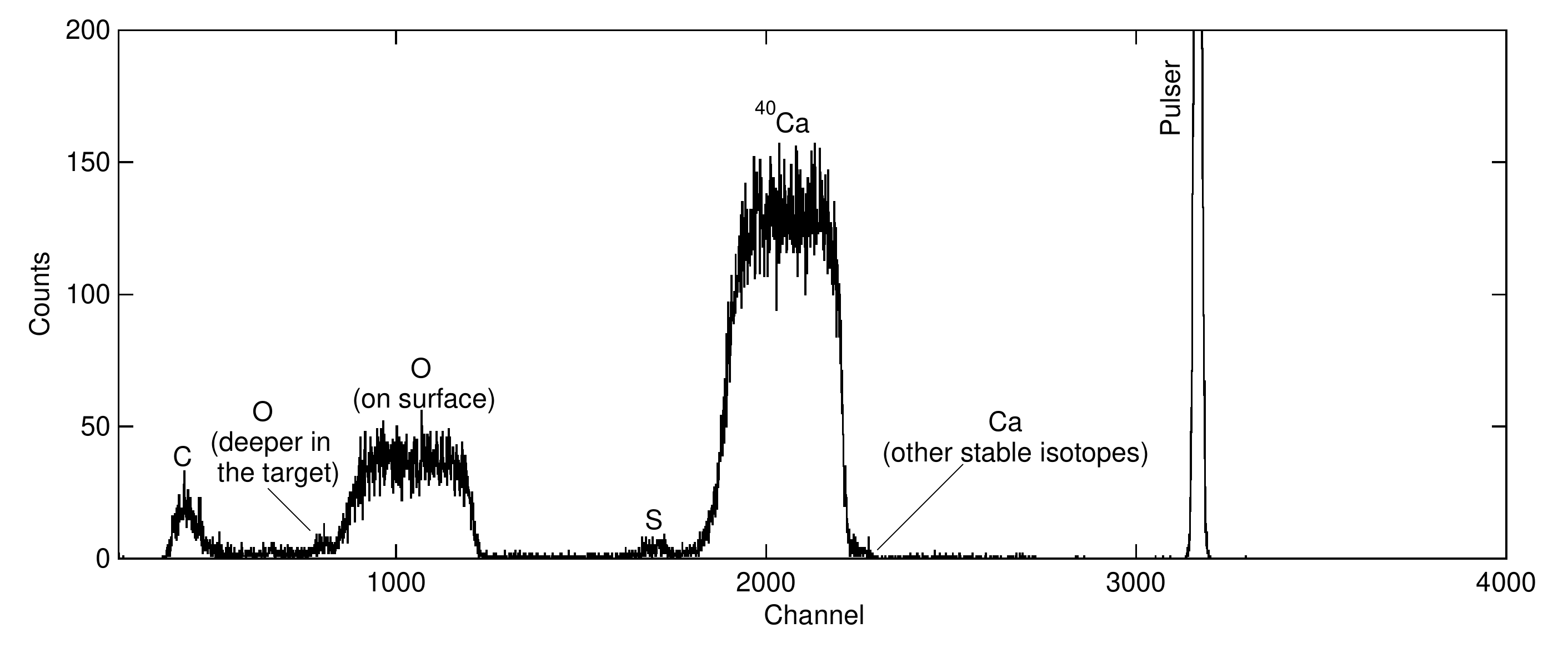}\\
\end{center}
\caption{\label{figure1}The spectrum from the RBS measurement on the calcium target. The peaks corresponding to the calcium and carbon contents of this target are shown. The wide oxygen peak comes from surface oxidation of calcium due to a few minutes exposure to air. The smaller oxygen peak is deeper within the target and could possibly be caused by floating the carbon foil in water or some oxygen contamination that preexisted on the carbon backing foil. The sulfur peak comes from residual CdS contaminations in the evaporator system.}
\end{figure*}

\noindent encouraged in Ref.~\cite{Lotay:2016} in order to search for potential unobserved low-spin proton resonances in the $^{39}$K $+$ $p$ system.\par
We performed a high-resolution charged-particle spectroscopy experiment via the $^{40}$Ca($^{3}$He, $\alpha$)$^{39}$Ca reaction. We specifically aimed to explore $E_{x}$($^{39}$Ca) $\sim$ 6 MeV region, where the energies of the astrophysically significant proton resonances are still ambiguous. This work presents the results.

\section{\label{Experiment}Experimental Setup and Data Analysis}

The experiment was performed at Triangle Universities Nuclear Laboratory (TUNL). A 21-MeV $^{3}$He$^{2+}$ beam ($\Delta\,E$/$E$ $\sim$ 3.5 $\times$ 10$^{-4}$) was delivered by the TUNL duoplasmatron ion source and 10-MV FN tandem Van de Graaff accelerator. The beam energy was analyzed using two high resolution 90$^{\circ}$ dipole magnets and focused to a spot size of 1-mm diameter on target. Typical beam intensity on target varied between 350 -- 1000 enA.\par
Three types of targets were employed: a calcium target for measuring the main reaction of interest ($^{40}$Ca($^{3}$He, $\alpha$)$^{39}$Ca); a silicon oxide target for calibration purposes; and a carbon target for background determination. The first two targets were separately produced by thermal evaporation of natural metallic calcium and SiO$_{2}$ powder onto a 38-$\mu$g/cm$^{2}$-thick natural carbon foil from the Arizona Carbon Foil Company~\cite{ACF-Metals}. A 38-$\mu$g/cm$^{2}$-thick foil comprised our carbon target.\par
Target thicknesses and stoichiometries were measured using Rutherford Backscattering Spectrometry (RBS) following the experiment. For the RBS, a 2-MeV $^{4}$He$^{2+}$ beam was employed using the same accelerator facility. The backscattered $\alpha$-particles were measured at 165$^{\circ}$ with respect to the beam axis using a single 100-$\mu$m-thick silicon surface barrier detector with 17-keV energy resolution. A gold target with a known thickness was used to calibrate the RBS spectra. During the RBS measurement, a pulser was used to adjust the gain of the silicon detector and monitor the electronics.\par
RBS revealed that i) the calcium target (see Fig.~\ref{figure1}) is composed of 58.3 $\mu$g/cm$^{2}$ Ca, 37.7 $\mu$g/cm$^{2}$ O, 37.8 $\mu$g/cm$^{2}$ C and 0.9 $\mu$g/cm$^{2}$ S, where the small sulfur contamination comes from residual CdS contaminations in the evaporator system; ii) the silicon oxide target is composed of 13.6 $\mu$g/cm$^{2}$ Si, 30.2 $\mu$g/cm$^{2}$ O, 36.8 $\mu$g/cm$^{2}$ C and 7.5 $\mu$g/cm$^{2}$ Ta, where the latter contamination comes from partial melting of the Ta evaporation boat towards the end of the evaporation; and iii) the carbon target is composed of 37.9 $\mu$g/cm$^{2}$ C. The uncertainties in these thicknesses were determined to be $\approx$10\%. This factor comes from a conservative estimation of the uncertainty of stopping powers of helium in calcium from SRIM~\cite{SRIM} where no experimental data are available.\par
The calcium target was fabricated at the beginning of the experiment and was exposed to air for less than 10 minutes when mounted into the main target chamber. Some degree of oxygen contamination was expected in the calcium target and the experiment was planned accordingly to avoid the states of interest in $^{39}$Ca being obscured by the oxygen contamination. No excited states from the sulfur and tantalum contaminations in the targets were observed.\par
The light reaction products were accepted by the TUNL high resolution Enge split-pole magnetic spectrograph~\cite{Setoodehnia:2016}, whose total solid angle acceptance was set to 1 msr for this experiment. The charged particles were dispersed by the spectrograph according to their momenta and were focused onto the spectrograph's focal plane. The magnetic field of the spec-

\begin{figure*}[ht]
 \begin{center}
  \subfloat{%
   \includegraphics[width=0.9\textwidth]{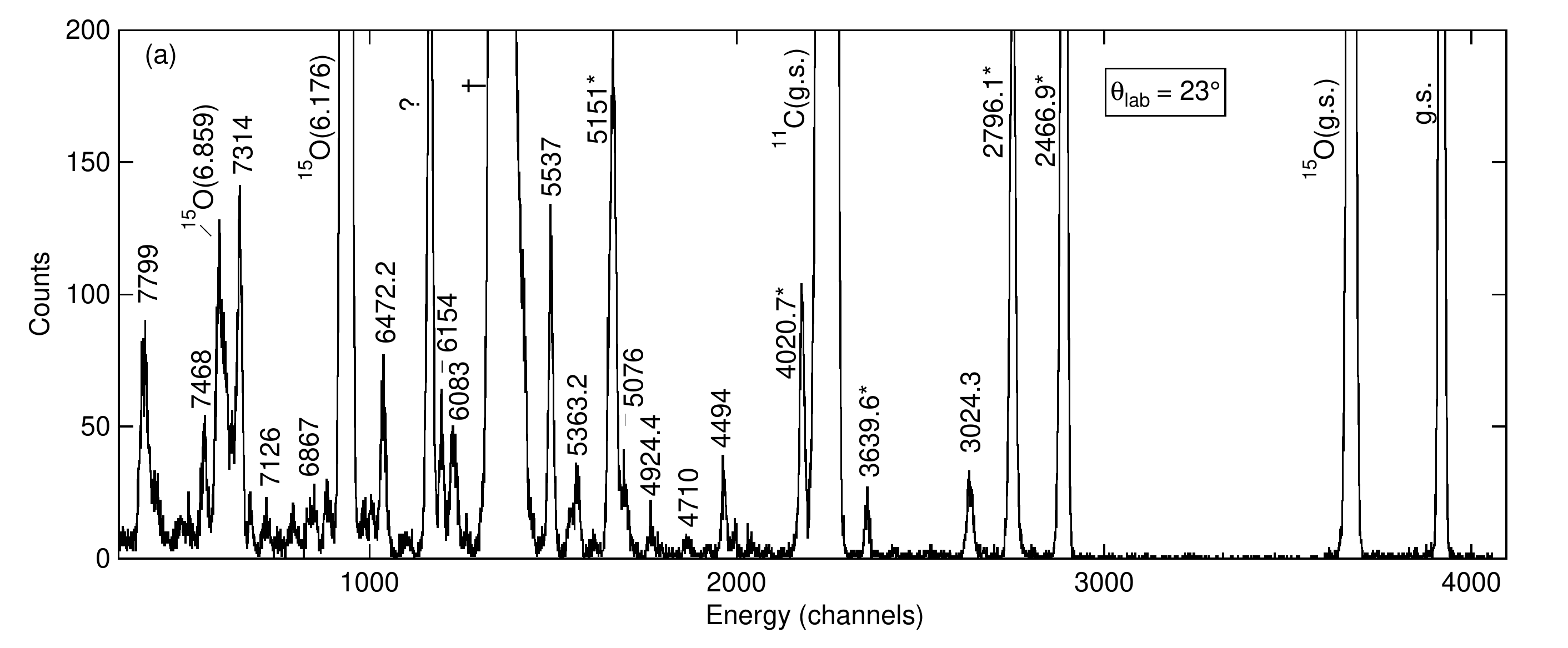}
   }\\
   \subfloat{%
   \includegraphics[width=0.9\textwidth]{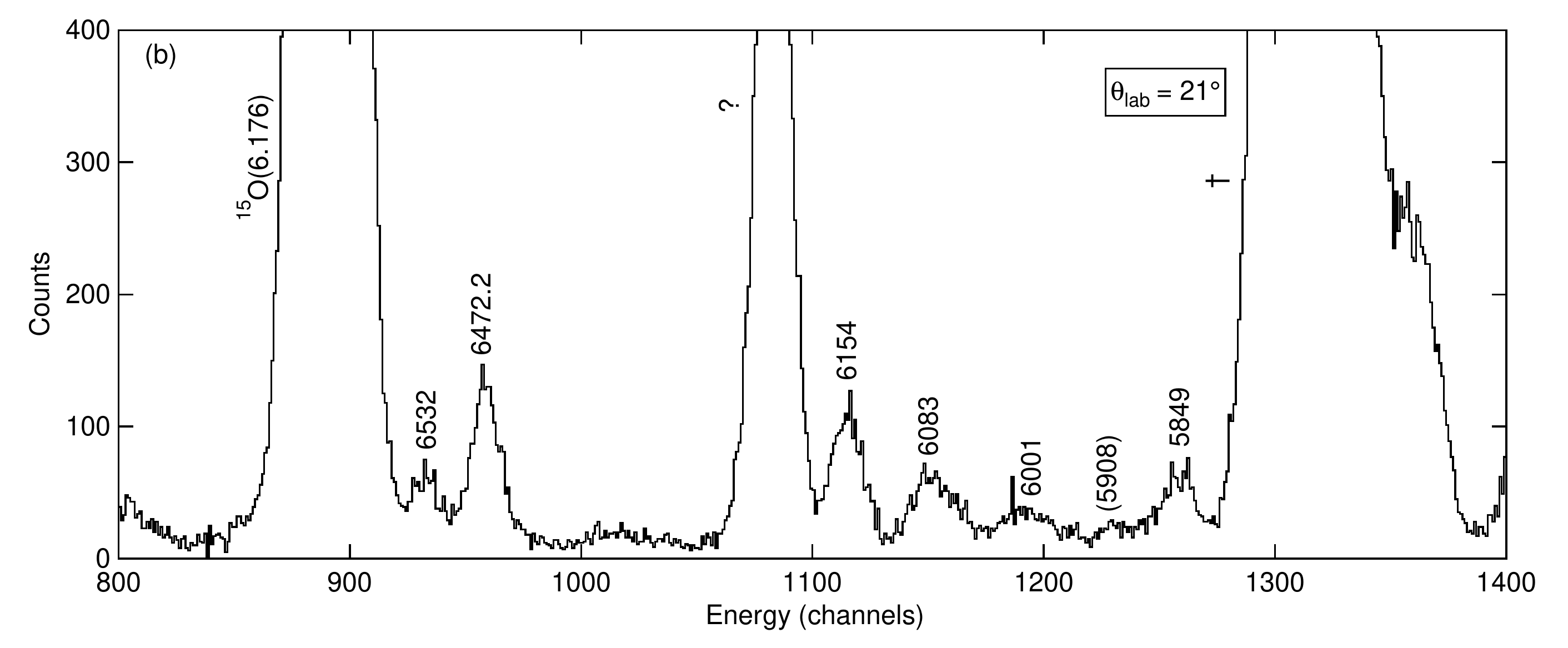}
   }\\
   \subfloat{%
   \includegraphics[width=0.9\textwidth]{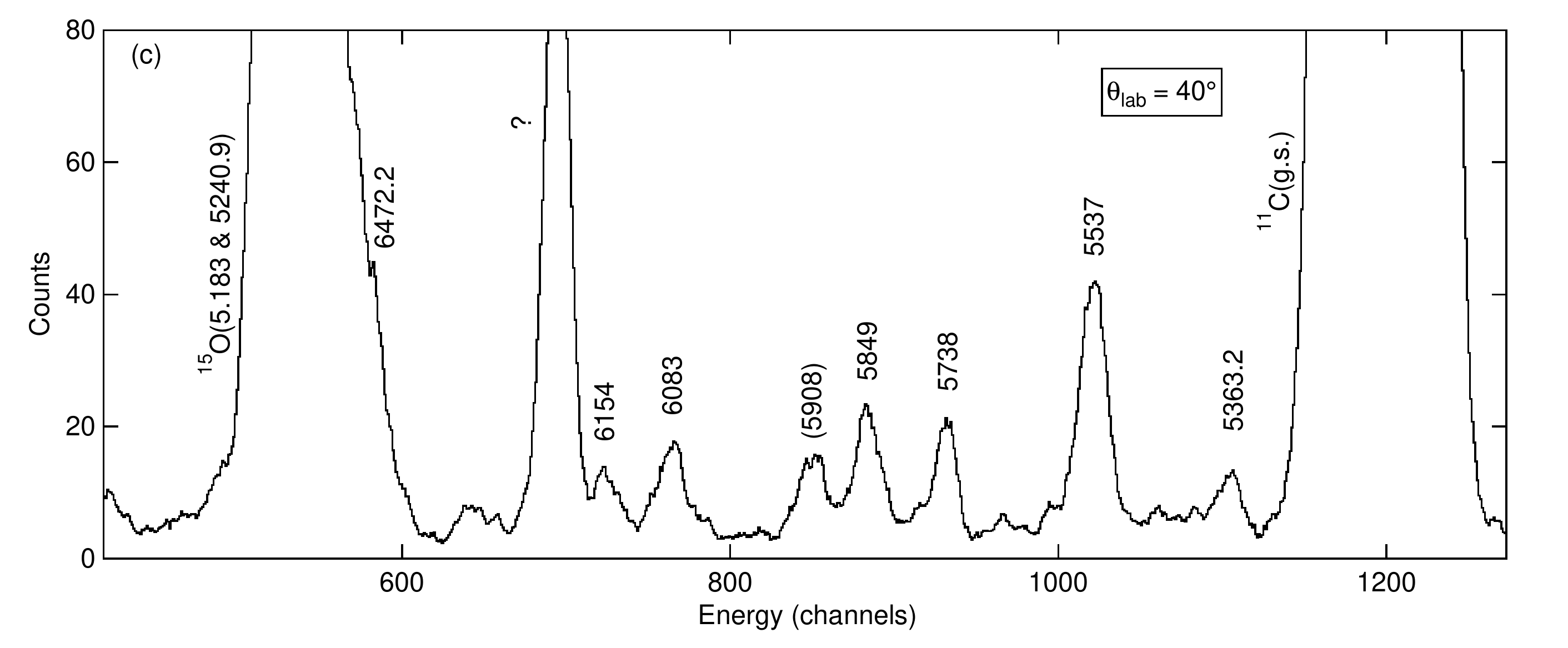}
   }
  \end{center}\vspace{-0.2cm}
\caption{\label{figure2}The spectra from the $^{40}$Ca($^{3}$He, $\alpha$)$^{39}$Ca reaction at $\theta_{lab}$ $=$ 23$^{\circ}$ (a), 21$^{\circ}$ (b) and 40$^{\circ}$ (c). At the latter two angles, the spectra are zoomed in on the region of astrophysical significance. Peaks corresponding to $^{39}$Ca states are labeled with energies (in keV) from the present work except those denoted by asterisks, which were used as internal calibration using energies from Ref.~\cite{Chen:2018}. For clarity, not all peaks are labeled. Ground states are indicated by g.~s. The main contaminant peaks are labeled with their parent nuclei and their energies (in MeV). The peak marked by $\dag$ consists of the 5183-keV and 5240.9-keV states of $^{15}$O, and the 2-MeV state of $^{11}$C. Tentative states are in parenthesis. For the discussion regarding to the peak denoted by ? mark, see \S~\ref{Energies}.}
\end{figure*}

\begin{figure*}[ht]
\begin{center}
\includegraphics[width=\textwidth]{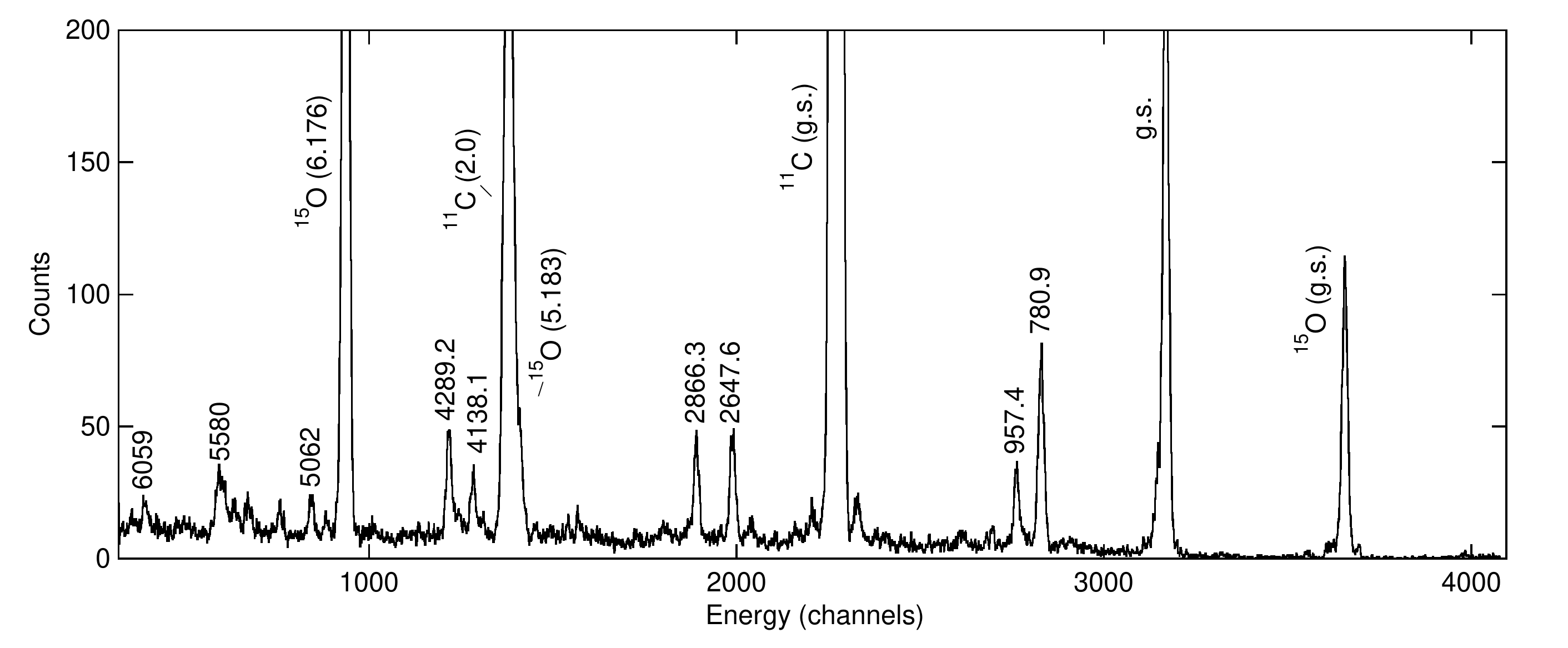}\\
\end{center}\vspace{-0.5cm}
\caption{\label{figure3}The spectrum from the $^{28}$Si($^{3}$He, $\alpha$)$^{27}$Si calibration reaction measured at $\theta_{lab}$ $=$ 19$^{\circ}$. Selected peaks corresponding to $^{27}$Si states are labeled with energies from Ref.~\cite{Basunia:2011} (in keV). For clarity, not all peaks are labeled. The main contaminant peaks are labeled with their parent nuclei and their energies (in MeV). g.~s.~indicates ground state.}
\end{figure*}

\noindent trograph was varied between 8.9 -- 9 kG so as to i) kinematically exclude the elastically and inelastically scattered $^{3}$He particles, and ii) to accept the $\alpha$-particles from the $^{40}$Ca($^{3}$He, $\alpha$) reaction whose radii of curvature laid on the focal plane between 68 cm to 84 cm, covered by the focal plane detector. These correspond to $^{39}$Ca excited states from ground state to 7799 keV.\par
To obtain optimal momentum resolution with the TUNL Enge split-pole spectrograph, a high resolution position sensitive focal plane detector was deployed. The detection system is described in detail elsewhere~\cite{Marshall:2018}. Identification of $\alpha$-particles was carried out by measuring their energy losses, residual energies and positions along the focal plane of the spectrograph. By placing software gates around the $\alpha$-particles, their momentum spectra were obtained at each spectrograph angle (see Fig.~\ref{figure2}). Peaks in these spectra represent the energy levels of $^{39}$Ca. The reaction products were measured at laboratory angles of 19${^\circ}$, 21${^\circ}$, 23${^\circ}$, 25${^\circ}$, 40${^\circ}$ and 44${^\circ}$. These scattering angles were particularly chosen to move the contaminant states away from the region of interest in $^{39}$Ca. Scattering angles lower than 19${^\circ}$ were not considered because the $^{15}$O state at 6176.3 keV would obscure the $^{39}$Ca region of interest.\par
The major contaminant peaks present in the spectra were the $E_{x}$ $=$ 0 -- 4804-keV states of $^{11}$C~\cite{Kelley:2012}, populated via the
$^{12}$C($^{3}$He, $\alpha$)$^{11}$C reaction, and the $E_{x}$ $=$ 0 -- 6859-keV states of $^{15}$O~\cite{Ajzenberg:1991}, populated via the
$^{16}$O($^{3}$He, $\alpha$)$^{15}$O reaction (see Fig.~\ref{figure2}).\par
The peaks observed in the spectra were fitted using a least-squares multi-Gaussian fit function to determine the peak centroids, widths,
and areas. The spectra were first calibrated using the known levels of $^{27}$Si~\cite{Basunia:2011} measured using the SiO$_{2}$ target (see Fig.~\ref{figure3}), and those of $^{15}$O~\cite{Ajzenberg:1991} produced by the ($^{3}$He, $\alpha$) reaction on the oxygen content of the calcium target. Once the well populated low-lying states of $^{39}$Ca were identified, they were used together with the strong $^{15}$O peaks and only a couple of statistically significant $^{27}$Si states (780.9 keV and 4289.2 keV) to recalibrate each $^{39}$Ca spectrum. The energies of the $^{39}$Ca internal calibration points were adopted from Ref.~\cite{Chen:2018} and are marked by asterisks in Table~\ref{tab:1}.\par
Calibration of each spectrum was performed using Bayesian framework explained in Ref.~\cite{Marshall:2018}. A cubic polynomial fit of the form $\rho$ $=$ $A\,+\,Bx\,+\,Cx^{2}\,+\,Dx^{3}$ was obtained according to this method for a set of calibration peaks at each angle. Here, $\rho$ and $x$ respectively correspond to the radius of curvature of an $\alpha$-particle traversing the spectrograph, and the channel number corresponding to the centroid of the calibration peak produced on the focal plane by that $\alpha$-particle. Coefficients $A$ to $D$ are fit parameters. These fits were used to derive the excitation energies for the unknown peaks in the spectrum at that angle.\par
Excitation energies' uncertainties reported in Table~\ref{tab:1} were computed from the statistical uncertainties in the corresponding peaks' centroids; uncertainties in the coefficients of the polynomial calibration fits; and the reproducibility of calibration peaks.\par
The mutually independent systematic uncertainties for the excitation energies at each angle (not included in Table~\ref{tab:1}) arise from: (1) $\pm$10\% uncertainties in the thicknesses of the calcium and SiO$_{2}$ targets affecting energy losses through these targets; (2) $\pm$7.3 keV uncertainty in the beam energy; and (3) the systematic uncertainties in the $Q$-values of the ($^{3}$He, $\alpha$) reactions on $^{16}$O, $^{28}$Si and $^{40}$Ca target nuclei, which are 0.49 keV, 0.11 keV, and 0.6 keV~\cite{Wang:2017}, respectively. The overall resultant systematic uncertainty in each excitation energy is 1.4

\begin{table*}[ht]
\caption{Weighted average (over all angles) excitation energies (in keV) of $^{39}$Ca from the present work in comparison with those measured in the selected previous work. Levels measured previously but not observed in the present work are omitted from the list. Ref.~\cite{Chen:2018} includes all previous experimental results for $^{39}$Ca. States used in the present work for internal energy calibration are denoted by an asterisk. The uncertainties reported for the present work do not include the $\pm$1.4 keV systematic uncertainty in our results.}
\begin{minipage}{\linewidth}
\renewcommand\thefootnote{\thempfootnote}
\setlength{\tabcolsep}{1.5pt}
\centering
\begin{tabular}{cccccccclcclc}
\toprule[1.0pt]\addlinespace[0.3mm] \multicolumn{2}{c}{$^{40}$Ca($^{3}$He, $\alpha$) Evaluation~\cite{JChen:2018}}\footnote{These are weighted averages between the results of previous $^{40}$Ca($^{3}$He, $\alpha$)$^{39}$Ca measurements of Refs.~\cite{Bock:1965,Yagi:1965,Hinds:1966,Lui:1980}.} & \multicolumn{1}{c}{\phantom{ab}} & \multicolumn{1}{c}{$^{40}$Ca($d, t$)~\cite{Doll:1976}} & \multicolumn{1}{c}{\phantom{ab}} & \multicolumn{2}{c}{$^{40}$Ca($p, d$)~\cite{Matoba:1993}} & \multicolumn{1}{c}{\phantom{ab}} & \multicolumn{2}{c}{$^{39}$Ca Evaluation~\cite{Chen:2018}} & \multicolumn{1}{c}{\phantom{ab}} & \multicolumn{2}{c}{($^{3}$He, $\alpha$) Present Work}\\
\cmidrule[0.05em]{1-2} \cmidrule[0.05em]{4-4} \cmidrule[0.05em]{6-7} \cmidrule[0.05em]{9-10} \cmidrule[0.05em]{12-13}\addlinespace[0.3mm]
\multicolumn{1}{c}{$E_{x}$ (keV)} & \multicolumn{1}{c}{$J^{\pi}$} & \multicolumn{1}{c}{} & \multicolumn{1}{c}{$E_{x}$ (keV)} & \multicolumn{1}{c}{} & \multicolumn{1}{l}{$E_{x}$ (keV)} & \multicolumn{1}{c}{$J^{\pi}$} & \multicolumn{1}{c}{} & \multicolumn{1}{l}{$E_{x}$ (keV)} & \multicolumn{1}{c}{$J^{\pi}$} & \multicolumn{1}{c}{} &
\multicolumn{1}{c}{$E_{x}$ (keV)} & \multicolumn{1}{c}{$J^{\pi}$}\\
\midrule[0.05em]\addlinespace[0.2mm]
0                     & 3/2$^{+}$ &  & 0        &  &  0        & 3/2$^{+}$   &  & 0          & 3/2$^{+}$                 &  & 0          & 3/2$^{+}$          \\
2473(10)              &           &  & 2470(15) &  &  2463(10) & 1/2$^{+}$   &  & 2466.9(5)  & 1/2$^{+}$                 &  & 2466.9(5)$\mbox{*}$ & 1/2$^{+}$ \\
2799(10)              &           &  & 2790(15) &  &  2791(10) & 7/2$^{-}$   &  & 2796.1(6)  & 7/2$^{-}$                 &  & 2796.1(6)$\mbox{*}$ & 7/2$^{-}$ \\
3032(10)              &           &  & 3030(15) &  &  3021(10) & 3/2$^{-}$   &  & 3025.1(9)  & 3/2$^{-}$                 &  & 3024.3(11) &  3/2$^{-}$ \\
3660(20)              &           &  & 3640(15) &  &  3636(10) &             &  & 3639.6(8)  & (9/2$^{-}$)               &  & 3639.6(8)$\mbox{*}$ &                     \\
3840(20)              &           &  & 3820(15) &  &  3820(10) &             &  & 3823.6(15) & (1/2,3/2,5/2)             &  & 3827.1(17) &                     \\
                      &           &  &          &  &  3886(10) &             &  & 3882(2)    & (3/2$^{-}$,5/2,7/2$^{+}$) &  & 3882(5)    &                     \\
                      &           &  & 3940(15) &  &  3943(10) &             &  & 3935.7(7)  & (3/2$^{-}$)               &  & 3936.2(22) &                     \\
4020(20)              &           &  & 4020(15) &  &  4016(10) & 1/2$^{+}$   &  & 4020.7(17) & 1/2$^{-}$)                &  & 4020.7(17)$\mbox{*}$ &                     \\
4320(20)              &           &  & 4320(15) &  &  4340(10) &             &  & 4332(10)   & (5/2,7/2)$^{-}$           &  & 4339(3)    &                     \\	
\hspace{1mm}4430?(20) &           &  & 4460(15) &  &  4432(10) & 5/2$^{+}$   &  & 4439(10)   & 3/2$^{+}$,5/2$^{+}$       &  & 4444(5)    &                     \\
4490(20)              &           &  &          &  &  4487(10) & 7/2$^{-}$   &  & 4488(10)   & 7/2$^{-}$                 &  & 4494(2)    &                     \\
4710(20)              &           &  &          &  &           &             &  & 4710(20)   &                           &  & 4710(3)    &                     \\
4920(20)              &           &  & 4940(15) &  &  4926(10) & (5/2$^{+}$) &  & 4929(10)   & 3/2$^{+}$,5/2$^{+}$       &  & 4924.4(24) &                     \\
5070(20)              &           &  &          &  &           &             &  & 5070(20)   & 3/2$^{+}$,5/2$^{+}$       &  & 5076(4)    &                     \\
5130(20)              & 5/2$^{+}$ &  & 5130(15) &  &  5128(10) & 5/2$^{+}$   &  & 5129(10)   & 5/2$^{+}$                 &  & 5116(4)    &            \\
		      &           &  &          &  &           &             &  & 5151(2)    & (11/2$^{-}$)              &  & 5151(2)$\mbox{*}$ &                     \\
                      &           &  &          &  &  5222(10) & 5/2$^{+}$   &  & 5222(10)   & 3/2$^{+}$,5/2$^{+}$       &  & 5223(4)    &                     \\
                      &           &  &          &  &  5364(10) &          	 &  & 5364(10)   &                       &  & 5363.2(16) &                     \\
                      &           &  &          &  &  5400(10) &          	 &  & 5400(10)   &                       &  & 5405(3)    &                     \\
		      &           &  &          &  &  5588(10) & (5/2$^{+}$) &  & 5588(10)   & 3/2$^{+}$,5/2$^{+}$       &  & 5537(6)    &                     3/2$^{+}$,5/2$^{+}$ \\
		      &           &  &          &  &  5673(10) & (5/2$^{+}$) &  & 5673(10)   & 3/2$^{+}$,5/2$^{+}$       &  & 5668(3)    &                     \\
5760(20)	      &           &  &          &  &  5720(10) & (7/2$^{-}$) &  & 5720(10)   & (5/2$^{-}$,7/2$^{-}$)     &  & 5738(3)    &                     \\ 	
                      &           &  &          &  &  5851(10) & 3/2$^{-}$   &  & 5851(10)   & 1/2$^{-}$,3/2$^{-}$       &  & 5849(3)    & \\
                      &           &  &          &  &           &             &  &            &                           &  & (5908(3))  &  \\
6000(20)              &           &  & 6010(15) &  &  6009(10) & (7/2$^{-}$) &  & 6008(10)   &                           &  & 6001(4)    &                     \\
		      &           &  &          &  &  6094(10) & (1/2$^{+}$) &  & 6094(10)   & (1/2$^{+}$)               &  & 6083(7)    & (7/2$^{+}$,9/2$^{+}$) \\
6150(20)              & 5/2$^{+}$ &  & 6160(15) &  &  6158(10) & 5/2$^{+}$   &  & 6157(10)   & 5/2$^{+}$                 &  & 6154(5)    & 3/2$^{+}$,5/2$^{+}$ \\
		      &		  &  &          &  &  6286(10) & 5/2$^{+}$   &  & 6286(10)   & 3/2$^{+}$,5/2$^{+}$       &  &   &  \\
		      &		  &  & 6450(30) &  &  6467(10) & 5/2$^{+}$	 &  & 6451(2)    & 3/2$^{+}$,5/2$^{+}$   &  & 6472.2(20) & (5/2$^{-}$,7/2$^{-}$) \\
		      &		  &  &          &  &  6514(10) & (5/2$^{+}$) &  & 6514(10)   & 3/2$^{+}$,5/2$^{+}$       &  & 6532(3)    & \\
		      &		  &  &          &  &  6580(10) & (7/2$^{-}$) &  & 6580(10)   & 5/2$^{-}$,7/2$^{-}$       &  & 6579(7)    &                     \\
		      &		  &  &          &  &  6629(10) &             &  & 6629(10)   &                           &  & 6645(7)    &                     \\
 	              &		  &  &          &  &  6794(10) & 5/2$^{+}$   &  & 6794(10)   & 3/2$^{+}$,5/2$^{+}$       &  & 6779(4)    &                     \\
 	              &           &  & 6820(30) &  &  6835(10) &             &  & 6834(10)   & 3/2$^{+}$,5/2$^{+}$       &  & 6867(6)    &                     \\
 	              &		  &  &          &  &  6954(10) &             &  & 6954(10)   & 5/2$^{-}$,7/2$^{-}$       &  & 6972(3)    &                     \\
 	              &		  &  &          &  &  7060(10) &             &  & 7060(10)   &                           &  & 7070(5)    &                     \\
 	              &		  &  &          &  &  7132(10) & 5/2$^{+}$   &  & 7132(10)   & 3/2$^{+}$,5/2$^{+}$       &  & 7126(9)    &                     \\
 	              &		  &  & 7210(30) &  &  7199(10) & 5/2$^{+}$   &  & 7199(10)   & 5/2$^{+}$                 &  & 7217(9)    &                     \\
 	              &		  &  &          &  &  7248(10) & 5/2$^{+}$   &  & 7248(10)   & 3/2$^{+}$,5/2$^{+}$       &  & 7240(9)    &                     \\
 	              &		  &  &          &  &  7310(10) &             &  & 7310(10)   & (5/2$^{-}$,7/2$^{-}$)     &  & 7314(6)    &                     \\
 	              &		  &  & 7380(30) &  &  7380(10) & 5/2$^{+}$   &  & 7380(10)   & 5/2$^{+}$                 &  & 7388(5)    &                     \\
 	              &		  &  &          &  &  7480(10) &             &  & 7480(10)   & (5/2$^{-}$,7/2$^{-}$)     &  & 7468(10)   &                     \\
 	              &		  &  &          &  &  7635(10) &             &  & 7635(10)   & (5/2$^{-}$,7/2$^{-}$)     &  & 7610(12)   &                     \\
 	              &		  &  & 7700(30) &  &  7711(10) & 5/2$^{+}$   &  & 7711(10)   & 3/2$^{+}$,5/2$^{+}$       &  & 7737(14)   &                     \\ 	
 	              &		  &  &          &  &  7773(10) &             &  & 7773(10)   &                           &  & 7799(15)   &                     \\[0.2ex]
\bottomrule[1.0pt]
\end{tabular}
\end{minipage}
\label{tab:1}
\end{table*}

\noindent keV. This should be added in quadrature to the uncertainties quoted in Table~\ref{tab:1}. To obtain the final $^{39}$Ca excitation energies, a weighted average was calculated (using V.AveLib utility code of Ref.~\cite{Birch:2012}) for each state over all the angles. The energy resolution defined as the peak Full Width at Half Maximum (FWHM) was 22 keV at 19$^{\circ}$ and 46 keV at 44$^{\circ}$. The loss of energy resolution at higher angles results from kinematic broadening~\cite{Marshall:2018,Enge:1958,Spencer:1967,Enge:1979} due to an increase in the energy straggling of the $\alpha$-particles through the target.

\section{\label{Results}Results}

\subsection{\label{Energies}Excitation energies of $^{39}$Ca}

45 states of $^{39}$Ca with excitation energies up to 7799 keV were observed in the present work and are listed in Table~\ref{tab:1}. Most of the measured energies in the present work are in agreement within 1 -- 2$\sigma$ with those measured in the previous $^{40}$Ca($^{3}$He, $\alpha$)$^{39}$Ca experiments~\cite{Bock:1965,Yagi:1965,Hinds:1966,Ronsin:1972,Lui:1980}, and with the excitation energies reported in the most recent evaluation of $^{39}$Ca excited states~\cite{Chen:2018}. The excitation energies above 7 MeV from the present work have larger uncertainties since these states lie far from the last calibration peak used, which is the $^{15}$O state at 6176.3(17) or 6859.4(9) keV~\cite{Ajzenberg:1991}, depending on the angle.\par
Three $J^{\pi}$ $=$ 5/2$^{+}$ excited states of $^{39}$Ca were identified previously~\cite{Iliadis:2002} to dominate the $^{38}$K($p, \gamma$)$^{39}$Ca reaction rate at temperatures characteristic of explosive hydrogen burning in novae. The excitation energies of these states were previously determined~\cite{Singh:2006} to be $E_{x}$ $=$ 6157(10), 6286(10) and 6460(10) keV. In the following, comparisons between the results of the present work with those of previous measurements will be described for these three states, as well as a few other states.\par
\textit{The 5537-keV level:} This state is one of the very few states whose energy from the present work is inconsistent with that measured previously~\cite{Matoba:1993}. In the present work, the 5537-keV state is observed at 4 angles. At 19$^{\circ}$ and 44$^{\circ}$, it has been obscured by the 5.183-MeV/5.2409-MeV states of $^{15}$O and by the ground state of $^{11}$C, respectively. The only other experiment in which this state has been observed in the past is the $^{40}$Ca($p, d$) measurement performed by Matoba {\em et al}.~\cite{Matoba:1993}. They observed a state at 5588 keV, which suffered from poor statistical significance in comparison with other observed states. It may be possible that since this state was populated weakly, it would have been a peak with rather large statistical uncertainty. However, Matoba {\em et al}.~\cite{Matoba:1993} did not assign uncertainties to their excitation energies. The 10-keV uncertainty assigned to their excitation energies in our Table~\ref{tab:1} comes from the decision made by nuclear data evaluators in Refs.~\cite{Singh:2006,Chen:2018} to assume a 10-keV uncertainty on all their excitation energies based on their energy resolution (25 -- 30 keV~\cite{Matoba:1993}).\par
\textit{The tentative 5908-keV level:} This state is not listed as an excited state of $^{39}$Ca in the most recent Evaluated Nuclear Structure Data File for $^{39}$Ca~\cite{Chen:2018}. The state follows expected systematics, but is weak at all 3 angles where it is observed (19$^{\circ}$, 40$^{\circ}$ and 44$^{\circ}$; see panel (c) in Fig.~\ref{figure2}). Hence, we tentatively assign it to $^{39}$Ca.\par
\textit{The 6154-keV level:} The excitation energy of 6154(5) keV measured in the present work agrees very well with all of the results of the previous measurements: 6150(20) keV~\cite{Lui:1980}, 6158(10) keV~\cite{Kozub:1968,Kallne:1975,Matoba:1993}, and 6160(15) keV~\cite{Doll:1976}.\par
\textit{The 6286-keV level:} At all 6 angles measured in the present work, a very strong peak (denoted by ? mark in Fig.~\ref{figure2}) is populated where the 6286-keV state, measured in Ref.~\cite{Matoba:1993}, was expected to be observed. If it is assumed that this peak belongs to $^{39}$Ca: (i) its excitation energy is derived to be 6226(10) keV, which agrees well with the $E_{x}$ $=$ 6200(50) keV measured in Ref.~\cite{Rapaport:1984} but is inconsistent with the $E_{x}$ $=$ 6286(10) keV from Ref.~\cite{Matoba:1993}. (ii) Its $\alpha$ angular distribution is in very good agreement with $J^{\pi}$ $=$ 3/2$^{+}$, 5/2$^{+}$ (see panel (h) of Fig.~\ref{figure6}), which is expected for the 6286-keV state~\cite{Matoba:1993}.\par
\begin{figure*}[ht]
 \begin{center}
   \includegraphics[width=0.7\textwidth]{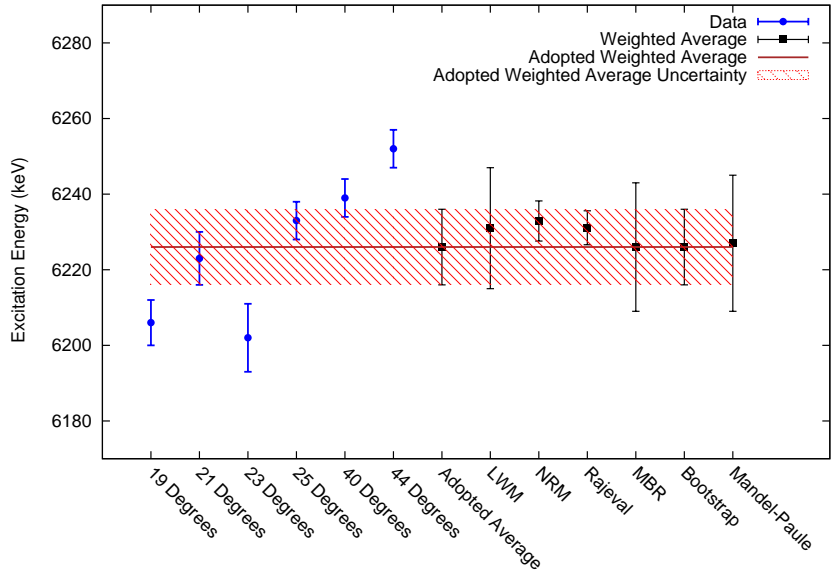}
 \end{center}\vspace{-4mm}
\caption{\label{figure4}(Color online) The individual calibrated energies (in keV, denoted by blue circles with error bars) at indicated $\theta_{lab}$ for the peak denoted by ? mark in Fig.~\ref{figure2} assuming it belongs to $^{39}$Ca. In the latter case, a weighted average energy of 6226(10) is adopted in the present work. The weighted average energies computed using different methods are also shown (black squares with error bars). The thick solid brown line is the adopted weighted average energy (from the bootstrap method). The red hatched band represents the uncertainty in the adopted weighted average energy.}
\end{figure*}

The $E_{x}$ $=$ 6226(10) keV result from the present work carries a large (relative to those of the other states) uncertainty of 10 keV because there is a rather large shift equal to 47 keV in the excitation energy of this state measured at each individual angle from 19$^{\circ}$ to 44$^{\circ}$ (see Fig.~\ref{figure4}). Even though this kinematic shift in energy is large and seems to be correlated with angle, there are no outlying data points using the criteria described in Ref.~\cite{Birch:2012}. Moreover, the present excitation energy was calculated using the bootstrap Monte Carlo method~\cite{Helene:2002} because it accounts for the data exhibiting large scatter between points.\par
Since this peak is statistically well populated at all angles observed, it was odd that its excitation energy should differ by about 47 keV from 19$^{\circ}$ to 44$^{\circ}$. It cannot belong to $^{11}$C or $^{15}$O since the expected states of these nuclei in this region of the spectra are all accounted for. The possibility of this peak belonging to $^{12}$C (from the $^{13}$C($^{3}$He, $\alpha$) reaction) or to other stable isotopes of calcium and sulfur (from contaminant isotopes in the target, see \S~\ref{Experiment}), fluorine and nitrogen (from potential beamline contaminants), chlorine and sodium (from potential contaminants in the water used to float the carbon backing foil of the calcium target) was examined. However, none of these cases produced an energy consistent with or even close to the known excitation energies of these nuclei. In addition, if the peak is assumed to belong to the contaminants considered above, the extracted excitation energy would exhibit a similar kinematic shift correlated with angle but of over four times more than the observed 47 keV. The shift in this state's extracted energy regardless of its assumed origin, however, means that we cannot rule out the possibility of having an unknown contaminant in our target. However, the RBS spectrum of Fig.~\ref{figure1} shows no evidence of a substantial amount of an unknown contaminant. Also, it seems less likely that a surface contamination of some kind which is small enough to remain undetected by the RBS measurement produces a peak which is statistically well populated at all angles. Furthermore, this peak is unlikely to be a doublet since its width is consistent with those of the known $^{39}$Ca single states in the spectrum at each angle. The contaminant peaks are usually wider due to kinematic broadening caused by their different reaction kinematics. Finally, since the individual calibrations at each angle produced results for other $^{39}$Ca states consistent with previously measured values, the possibility of problems with the present calibrations are confidently ruled out.\par
Considering all the evidence presented here, and to be conservative, we have claimed the peak denoted by ? mark in Fig.~\ref{figure2} as an unidentified peak.\par

\textit{The 6472.2-keV level:} Our measured excitation energy of 6472.2(20) keV does not agree within 2$\sigma$ with $E_{x}$ $=$ 6450(2) keV~\cite{Lotay:2016,Christian:2018}. The latter measurement was carried out with three different beam energies, including 27.17 MeV. This value was chosen to cover the center-of-mass energy range of 689 $\pm$ 13 keV across the volume of the DRAGON gas target~\cite{Lotay:2016} and they observed a resonance at 679(2) keV (corresponding to $E_{x}$ $=$ 6450(2) keV), which was located downstream the center of the gas target. It seems unlikely for the 679-keV resonance to be the same state as the 701.3-keV resonance, which is the center-of-mass energy equivalent to the 6472.2-keV state observed in the present work. The 701.3(25)-keV\footnote{The uncertainty in the resonance energy comes from the quadratic sum of $\Delta$$E_{x}$ for the 6472.2-keV state, the uncertainty in the $Q$-value of the $^{38}$K($p, \gamma$) reaction, and the present $\pm$1.4-keV systematic uncertainty in the excitation energy.} resonance is within the center-of-mass energy window of the measurement of Refs.~\cite{Lotay:2016,Christian:2018}. However, there

\begin{table*}[ht]
\begin{threeparttable}[b]
\caption{\label{tab:2}Global optical potential model parameters for the present DWBA analysis. The potential depths were varied to reproduce the correct binding energies corresponding to each channel. For neutron binding potential, see text.}
\centering
\setlength{\tabcolsep}{4pt} %adds space between columns in the table
\begin{tabular}{clllllllllllllll}
\toprule[1.0pt]\addlinespace[0.3mm] Reaction  & $V_{R}$ & $r_{R}$ & $a_{R}$ & $V_{I}$ & $r_{I}$ & $a_{I}$ & $W_{D}$ & $r_{D}$ & $a_{D}$ & $V_{so}$ & $r_{so}$ & $a_{so}$ & $r_{0c}$ & $E_{beam}^{lab}$ \\
Channel & (MeV) & (fm) & (fm) & (MeV) & (fm) & (fm) & (MeV) & (fm) & (fm) & (MeV) & (fm) & (fm) & (fm) & (MeV) \\ \hline\hline\addlinespace[0.5mm]
$^{3}$He $+$ $^{40}$Ca\tnote{a} & 148.33 & 1.2 & 0.72 & 34.77 & 1.4 & 0.88 &  &  &  & 2.5 & 1.2 & 0.72 & 1.3 & 21 \\
$^{4}$He $+$ $^{39}$Ca\tnote{b} & 160.9 & 1.3421 & 0.6578 & 0 & 1.4259 & 0.5587 & 25.9079 & 1.2928 & 0.6359 & 0 & 1.2686 & 0.85 & 1.35 & 21 \\[0.2ex]
\bottomrule[1.0pt]
\end{tabular}
\begin{tablenotes}
\item[a] Adopted from Ref.~\cite{Perey:1976}.
\item[b] Adopted from Ref.~\cite{Su:2015}.
\end{tablenotes}
\end{threeparttable}
\end{table*}

\noindent are two issues here: (a) the 701.3-keV resonance would have been placed at the very beginning of the DRAGON extended gas target, where the gas density is not homogenized or even optimized due to differential pumping system. This is clearly observed from panel (b) of Fig.~3 in Ref.~\cite{Christian:2018}. (b) Even if we do not consider the dramatic drop in the gas density, and the resultant change in the stopping powers, where the 701.3-keV resonance would have been located, the DRAGON recoil acceptance is usually set for a resonance near the center of the gas target~\cite{Engel:2003}. So, the measurement of Refs.~\cite{Lotay:2016,Christian:2018} would most likely be insensitive to recoils from the 701.3-keV resonance located at the beginning of the DRAGON gas target. Therefore, it is not surprising that the 701.3-keV resonance was not observed in the measurement of Refs.~\cite{Lotay:2016,Christian:2018}. The present result for $E_{x}$ $=$ 6472.2(20) keV agrees with $E_{x}$ $=$ 6467(10) keV~\cite{Matoba:1993} and 6450(30) keV~\cite{Doll:1976}.

\subsection{\label{Spin-Parities}Spin-parities of $^{39}$Ca excited states}

To determine the spin-parity values of the $^{39}$Ca states observed in the present work, the differential cross sections in the laboratory system were obtained from the procedure presented in Ref.~\cite{Setoodehnia:2011}. The measured differential cross sections in the laboratory system were converted to those in the center-of-mass system following the formalism presented in Appendix C of Ref.~\cite{Iliadis:2007}.\par
The theoretical angular distributions of the ($^{3}$He, $\alpha$) cross sections were also computed via Distorted-Wave Born Approximation (DWBA) calculations using the one-step finite-range transfer formalism described in Ref.~\cite{Thompson:2009}. DWBA calculations were performed using FRESCO~\cite{Thompson:1988}.\par
The distorted waves in the entrance and exit channels were calculated using global optical interaction potentials given in Table~\ref{tab:2}.\par
Following Ref.~\cite{Denikin:2015}, the $\alpha$-particle wave functions were computed from the $^{3}$He $+$ $n$ interaction assuming a Gaussian
potential of the form:
\begin{equation}
V(r)\,=\,-V_{G}\exp\Big(-\frac{r^{2}}{R_{G}^{2}}\Big),
\end{equation}
\noindent where $R_{G}$ $=$ 2.452 fm~\cite{Denikin:2015}, while the potential depth $V_{G}$ was varied to reproduce the $S_{n}$($^{4}$He) $=$ –20.5776 MeV~\cite{Wang:2017}, where $S_{n}$ is the neutron separation energy.\par
To describe the interaction of $^{39}$Ca $+$ $n$ $\rightarrow$ $^{40}$Ca, a volume Woods-Saxon potential of the form:
\begin{equation}
V(r)\,=\,\frac{V_{R}}{1\,+\,\exp\Big(\frac{r\,-\,r_{0R}\,A_{T}^{1/3}}{a_{R}}\Big)},
\end{equation}

\noindent where index $R$ refers to the real part of the potential, $r_{0R}$ was considered to be equal to 1.25 fm for the present analysis, $a_{R}$ is the diffuseness parameter set equal to 0.65 fm here, and $A_{T}$ is the mass of target (equal to 39 for $^{39}$Ca). The real depth of the potential is described by $V_{R}$ and was varied to reproduce the correct values of neutron binding energies of 15635.0(6) keV~\cite{Wang:2017} in the $^{40}$Ca nucleus when both $^{39}$Ca and $^{40}$Ca are in their ground states, and 15635 $+$ $E_{x}$($^{39}$Ca) when $^{39}$Ca is in an excited state with energy $E_{x}$ (in keV).\par
The theoretical angular distribution curves obtained from FRESCO were normalized to the center-of-mass experimental differential cross sections using linear fits with zero intercepts. Finally, DWBA calculations were performed for the first four bound states (including the ground state) of $^{39}$Ca with known spin-parity to confirm the validity of the optical potential models used. The $J^{\pi}$ values obtained from the present work for these cases agree with those found in Ref.~\cite{Chen:2018}. We have also investigated the spin-parity of the present 5537-keV state to be more confident that we can assume it may be the same state as the 5588-keV state observed by Matoba {\em et al}.~\cite{Matoba:1993}.\par
\begin{figure*}[ht]
\begin{center}
\includegraphics[width=\linewidth]{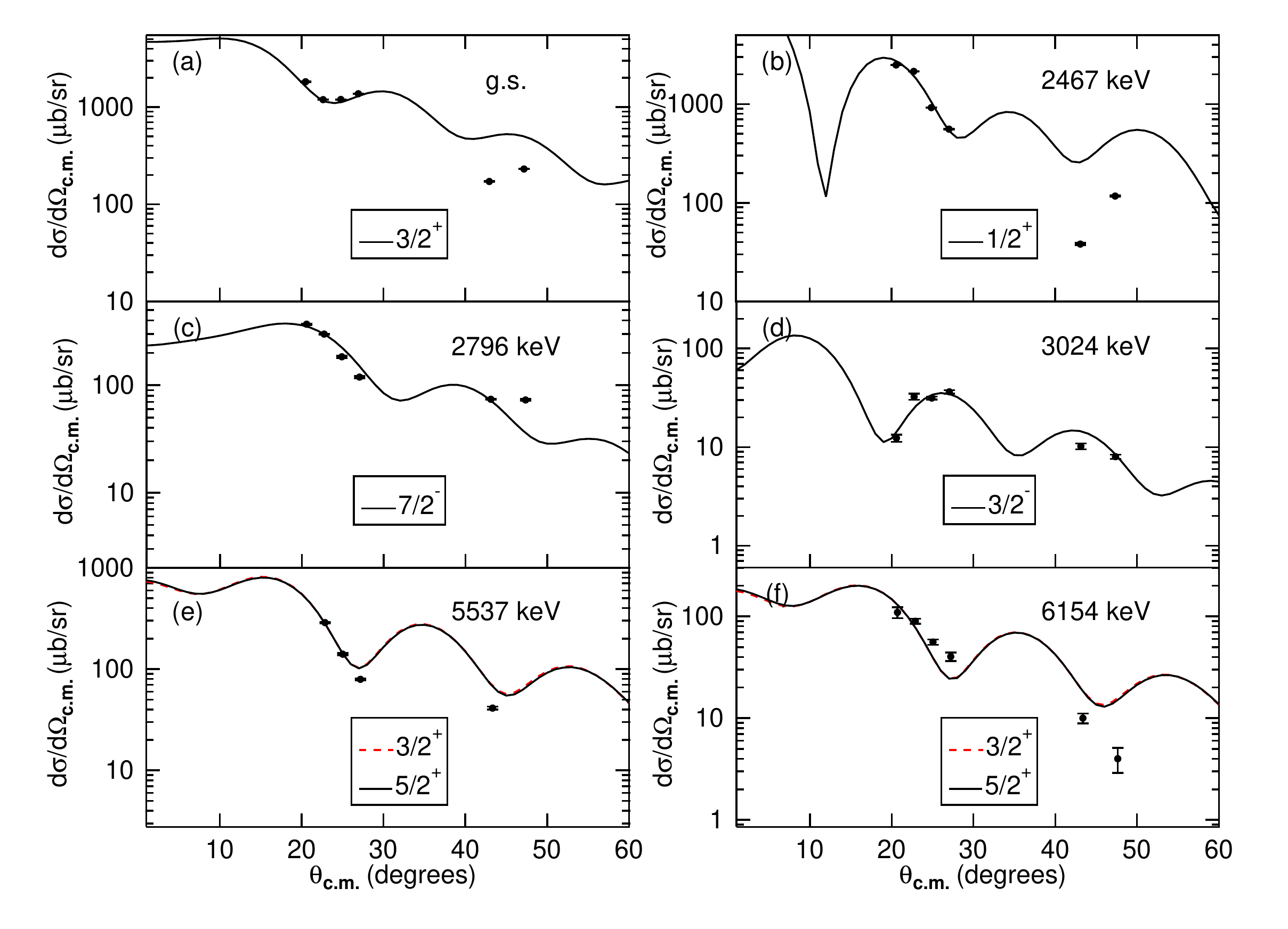}\\
\end{center}
\vspace{-1.2cm}
\caption{\label{figure5}(Color online) $\alpha$ angular distributions populating states of $^{39}$Ca compared with the DWBA curves (in black or red) calculated using FRESCO~\cite{Thompson:1988}. The filled circles with error bars are the measured differential cross sections (in the center-of-mass system) of the $\alpha$-particles from the $^{40}$Ca($^{3}$He, $\alpha$)$^{39}$Ca reaction. If not shown, the error bar is smaller than the point size.}
\end{figure*}
\begin{figure*}[ht]
\begin{center}
\includegraphics[width=\linewidth]{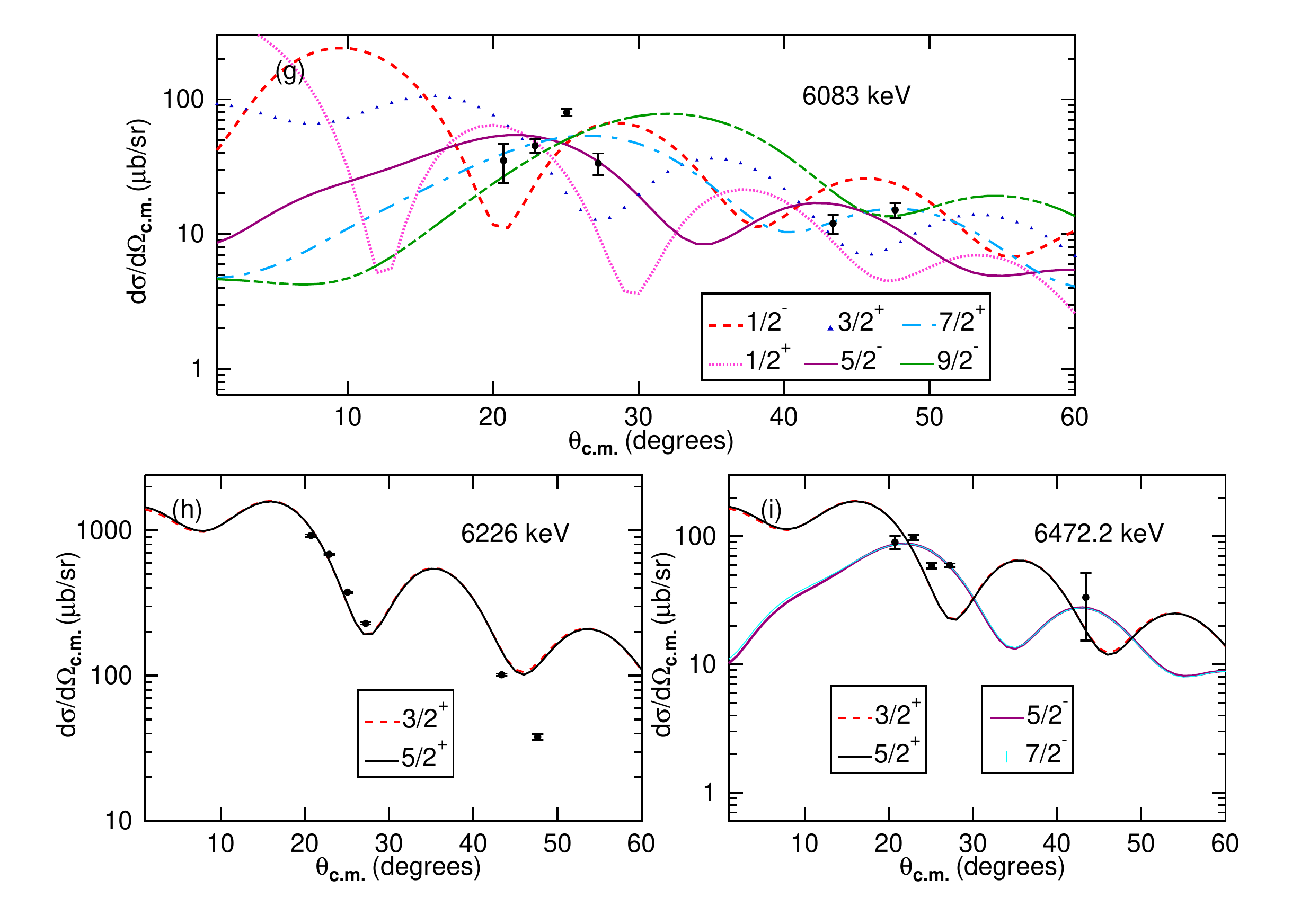}\\
\end{center}
\vspace{-1.2cm}
\caption{\label{figure6}(Color online) $\alpha$ angular distributions populating states of $^{39}$Ca compared with the DWBA curves (in black or color) calculated using FRESCO~\cite{Thompson:1988}. The filled circles with error bars are the measured differential cross sections (in the center-of-mass system) of the $\alpha$-particles from the $^{40}$Ca($^{3}$He, $\alpha$)$^{39}$Ca reaction. If not shown, the error bar is smaller than the point size. Panel (h) refers to the peak denoted by ? mark in Fig.~\ref{figure2} under the assumption that it belongs to $^{39}$Ca.}
\end{figure*}

Figures~\ref{figure5} and~\ref{figure6} show the present $\alpha$ angular distribution plots for the first four $^{39}$Ca bound states, the 5537-keV state,

\noindent as well as those $^{39}$Ca proton resonances (observed in at least 5 angles), which lie within the energy window important for determination of the $^{38}$K($p, \gamma$)$^{39}$Ca reaction rate at the nova temperature regime (see \S~\ref{Rate}). In what follows, we briefly compare the spin-parities derived in the present work with those of the previous measurements. Considering that neither our beam nor our target was polarized, \textit{l} $+$ \textit{s} and \textit{l} $-$ \textit{s} transitions, where \textit{l} is the orbital angular momentum and $s$ $=$ 1/2 is the spin of the transferred neutron, could not be differentiated.\par
The first 4 bound states: The spin-parities of these states are already known, and our angular distribution plots agree with the known assignments (see panels (a) to (d) of Fig.~\ref{figure5}). For the ground and first excited states of $^{39}$Ca, our theoretical DWBA fits are much higher than the data at large angles (see panels (a) and (b) of Fig.~\ref{figure5}). This is a common feature of transfer measurements, where the DWBA approximation is known to poorly reproduce large-angle cross sections.\par
The 5537-keV state: Even though the energy of this state is inconsistent with that previously measured~\cite{Matoba:1993} (see Table~\ref{tab:1}), the spin-parity assignments of $J^{\pi}$ $=$ 3/2$^{+}$,5/2$^{+}$ obtained in Ref.~\cite{Matoba:1993} agree well with our angular distribution data for this state (see panel (e) of Fig.~\ref{figure5}).\par
The 6083-keV state: A tentative $J^{\pi}$ $=$ (1/2$^{+}$) assignment is associated with this state~\cite{Matoba:1993}. We have obtained DWBA fits (see panel (g) of Fig.~\ref{figure6}) for $J^{\pi}$ $=$ 1/2$^{+}$, 1/2$^{-}$ (identical to 3/2$^{-}$), 3/2$^{+}$ (identical to 5/2$^{+}$), 5/2$^{-}$ (identical to 7/2$^{-}$), 7/2$^{+}$ (identical to 9/2$^{+}$), and 9/2$^{-}$ (identical to 11/2$^{-}$). Out of all these assignments, the best fit is obtained for $J^{\pi}$ $=$ 7/2$^{+}$, 9/2$^{+}$. These two identical fits give the minimum $\chi^{2}$/$\nu$ $=$ 8.8. The $\chi^{2}$/$\nu$ for $J^{\pi}$ $=$ 1/2$^{+}$ is a factor of 4 larger. Thus, we have assigned a tentative $J^{\pi}$ $=$ (7/2$^{+}$, 9/2$^{+}$) to this state. It should be noted that the uncertainties on the cross sections provided in Figs.~\ref{figure5} and~\ref{figure6} are only based on statistical uncertainties on the areas of the peaks, which are rather small. Furthermore, DWBA fits rarely pass through all the data points and they are dominated by uncertainties in the optical model; thus, it is unlikely to obtain a $\chi^{2}$/$\nu$ close to one without these contributions.\par
\textit{The 6154-keV state}: This state is known to be a 5/2$^{+}$ state~\cite{Chen:2018}. This assignment agrees well with the present $\alpha$ angular distribution data at lower angles (see panel (f) of Fig.~\ref{figure5}). However, our theoretical DWBA calculations are higher than the experimental data at larger angles. Also, we cannot distinguish between 5/2$^{+}$ and 3/2$^{+}$ assignments.\par
\textit{The 6472.2-keV state}: The spin-parity assignment of this state is determined as $J^{\pi}$ $=$ 3/2$^{+}$, 5/2$^{+}$ based on the measurements of Refs.~\cite{Matoba:1993,Doll:1976}. These two assignments cannot be immediately rejected from the present $\alpha$ angular distribution data based on visual inspection of the fits; however, much better fits are obtained with $J^{\pi}$ $=$ 5/2$^{-}$, 7/2$^{-}$ (see panel (i) of Fig.~\ref{figure6}). The $\chi^{2}$/$\nu$'s of our fits are as follows: $J^{\pi}$ $=$ 3/2$^{+}$, 5/2$^{+}$: $\chi^{2}$/$\nu$ $=$ 100; and $J^{\pi}$ $=$ 5/2$^{-}$, 7/2$^{-}$: $\chi^{2}$/$\nu$ $=$ 10. $J^{\pi}$ $=$ 1/2$^{\pm}$, 3/2$^{-}$, 7/2$^{+}$, 9/2$^{\pm}$ and 11/2$^{\pm}$ were also fitted to the present $\alpha$ angular distribution data. But they all resulted in having $\chi^{2}$/$\nu$ $\gg$ 10. We have therefore adopted a tentative assignment of (5/2$^{-}$, 7/2$^{-}$) for this state. Considering that the present assignment differs from that assumed for the 6450(2)-keV state of Refs.~\cite{Lotay:2016,Christian:2018}, it is less likely that these two states are the same.\par
A future measurement of the $\alpha$ angular distributions from the $^{40}$Ca($^{3}$He, $\alpha$) reaction at laboratory angles lower than 19$^{\circ}$ could help make definite conclusions on spin/parity assignments of the 6083- and 6472.2-keV states.

\subsection{\label{Rate}The $^{38}$K($p, \gamma$)$^{39}$Ca reaction rate}

Over the temperatures characteristic of explosive hydrogen burning in novae, the Gamow window~\cite{Iliadis:2007} for the $^{38}$K($p, \gamma$)$^{39}$Ca reaction ($Q$ $=$ 5770.92(63) keV~\cite{Wang:2017}) spans $E_{cm}$ $=$ 140 -- 615 keV. Therefore, the $^{38}$K($p, \gamma$)$^{39}$Ca reaction rate is dominated by contributions from isolated and narrow $^{38}$K $+$ $p$ resonances corresponding to $^{39}$Ca excited states with 5911 keV $\lessapprox$ $E_{x}$ $\lessapprox$ 6386 keV. In particular, the largest impact on the $^{38}$K($p, \gamma$)$^{39}$Ca reaction rate comes from those excited states within this energy range with $J^{\pi}$ $=$ 5/2$^{+}$ and 7/2$^{+}$, as such states correspond to \textit{l} $=$ 0 proton captures on the ground state of $^{38}$K with $J^{\pi}$ $=$ 3$^{+}$.\par
Table~\ref{tab:1} shows that there are 5 states in the region between 5911 keV $\lessapprox$ $E_{x}$ $\lessapprox$ 6386 keV. The 5908-keV and 6001-keV states have unknown spins-parities. The 6083-keV state has a tentative 7/2$^{+}$,9/2$^{+}$ assignment, and the 6154-keV and 6286-keV~\cite{Matoba:1993} states have $J^{\pi}$ $=$ 3/2$^{+}$,5/2$^{+}$ assignments. These states are candidates for dominating the $^{38}$K($p, \gamma$)$^{39}$Ca reaction rate over nova temperature regime.\par
The existence of the 5908-keV state is tentative. Furtherm-

\begin{table*}[ht]
\begin{threeparttable}[ht]
\caption{\label{tab:3}$^{39}$Ca level parameters for the $^{38}$K($p, \gamma$)$^{39}$Ca resonant reaction rate. The $\pm$1.4-keV systematic uncertainty in the present work is added in quadrature to the uncertainties in $E_{x}$ given in Table~\ref{tab:1}.}
\centering
\setlength{\tabcolsep}{5pt} %adds space between columns in the table
\begin{tabular}{lllllllll}
\toprule[1.0pt]\addlinespace[0.3mm] \multicolumn{4}{c}{from Ref.~\cite{Christian:2018}} & \multicolumn{1}{c}{\phantom{ab}} & \multicolumn{4}{c}{Present Work}\\
\cmidrule[0.05em]{1-4} \cmidrule[0.05em]{6-9}\addlinespace[0.3mm]
\multicolumn{1}{l}{$E_{x}$ (keV)} & \multicolumn{1}{l}{$E_{r}$ (keV)} & \multicolumn{1}{l}{$\omega\gamma$ (meV)} & \multicolumn{1}{l}{$J^{\pi}$} & \multicolumn{1}{l}{} & \multicolumn{1}{l}{$E_{x}$ (keV)} & \multicolumn{1}{l}{$E_{r}$ (keV)} & \multicolumn{1}{l}{$\omega\gamma$ (meV)} & \multicolumn{1}{l}{$J^{\pi}$}\\
\midrule[0.05em]\addlinespace[0.2mm]
6157(10)\tnote{a} & 386(10) &  $\leq$ 2.54\tnote{b} & 5/2$^{+}$ &  & 6154(5) & 383(5) & $\leq$ 2.6 & 5/2$^{+}$ \\
6286(10)\tnote{a} & 515(10) & $\leq$ 18.4\tnote{b} & 5/2$^{+}$ &  & 6286(10)\tnote{a} & 515(10) & $\leq$ 18.4\tnote{b} & 5/2$^{+}$ \\
6450(2)\tnote{c} & 679(2)\tnote{d} & 120(25)\tnote{d} & 5/2$^{+}$ &  & 6472.2(24) & 701.3(25) & 126(39) & (5/2$^{-}$) \\[0.2ex]
\bottomrule[1.0pt]
\end{tabular}
\begin{tablenotes}
\item[a] Adopted from Ref.~\cite{Singh:2006}. Resonance was not observed.
\item[b] This value is the upper limit at 90\% confidence level~\cite{Christian:2018}.
\item[c] Derived from the measured resonance energy~\cite{Lotay:2016,Christian:2018}.
\item[d] See text in \S~\ref{Rate} regarding the uncertainty.
\end{tablenotes}
\end{threeparttable}
\end{table*}

\noindent ore, nothing is known about the decay schemes, $\gamma$ branching ratios, lifetimes, or proton spectroscopic factors of the 5908-keV, 6001-keV, and 6083-keV levels. Moreover, mirror and/or isobaric analog levels of these states are also unknown. Therefore, their proton and gamma widths, and thus resonance strengths, cannot be reliably estimated. In conclusion, we do not have sufficient information to make educated guesses about the properties of these three proton unbound states in $^{39}$Ca.\par
We have therefore followed Refs.~\cite{Lotay:2016,Christian:2018} and have calculated the contributions of the two remaining levels (at 6154(5) keV from the present work and 6286(10) keV observed in Ref.~\cite{Matoba:1993}) to the $^{38}$K($p, \gamma$)$^{39}$Ca reaction rate. Furthermore, to be consistent with Refs.~\cite{Lotay:2016,Christian:2018} for reaction rate calculation, we will similarly also included the higher energy resonance corresponding to the present 6472.2(24)\footnote{The $\pm$1.4-keV systematic uncertainty in the present work is added in quadrature to the uncertainty given in Table~\ref{tab:1}.}-keV state in the calculation of the $^{38}$K($p, \gamma$)$^{39}$Ca reaction rate even though it falls outside the range of interest in $^{39}$Ca over the nova temperature regime The resonance parameters for these states are given in Table~\ref{tab:3}.\par
The strengths of the 383-keV and 701.3-keV resonances from the present work were scaled from those provided in Ref.~\cite{Christian:2018}. The latter study has used the method of thick target yield curve to compute resonance strength from~\cite{Christian:2018}:
\begin{equation}
\omega\gamma\,=\,\frac{2N_{r}\epsilon}{N_{b}\eta\lambda^{2}},
\end{equation}

\noindent where $N_{r}$ and $N_{b}$ are number of recoils and beam particles, respectively; $\epsilon$ is center-of-mass stopping power; $\eta$ is heavy ion detection efficiency; and $\lambda$ is the center-of-mass de Broglie wavelength.\par
In the above formula, $\epsilon$, $\eta$ and $\lambda$ are the energy dependent factors. Ref.~\cite{Christian:2018} has provided stopping powers for three different beam energies in the laboratory system. These were used to obtain a linear fit between $\epsilon$ and energy. Heavy ion detection efficiency is only given for one beam energy corres-

\begin{figure*}[ht]
 \begin{center}
  \subfloat[Individual resonance contribution to the reaction rate.]{%
   \includegraphics[width=0.5\textwidth]{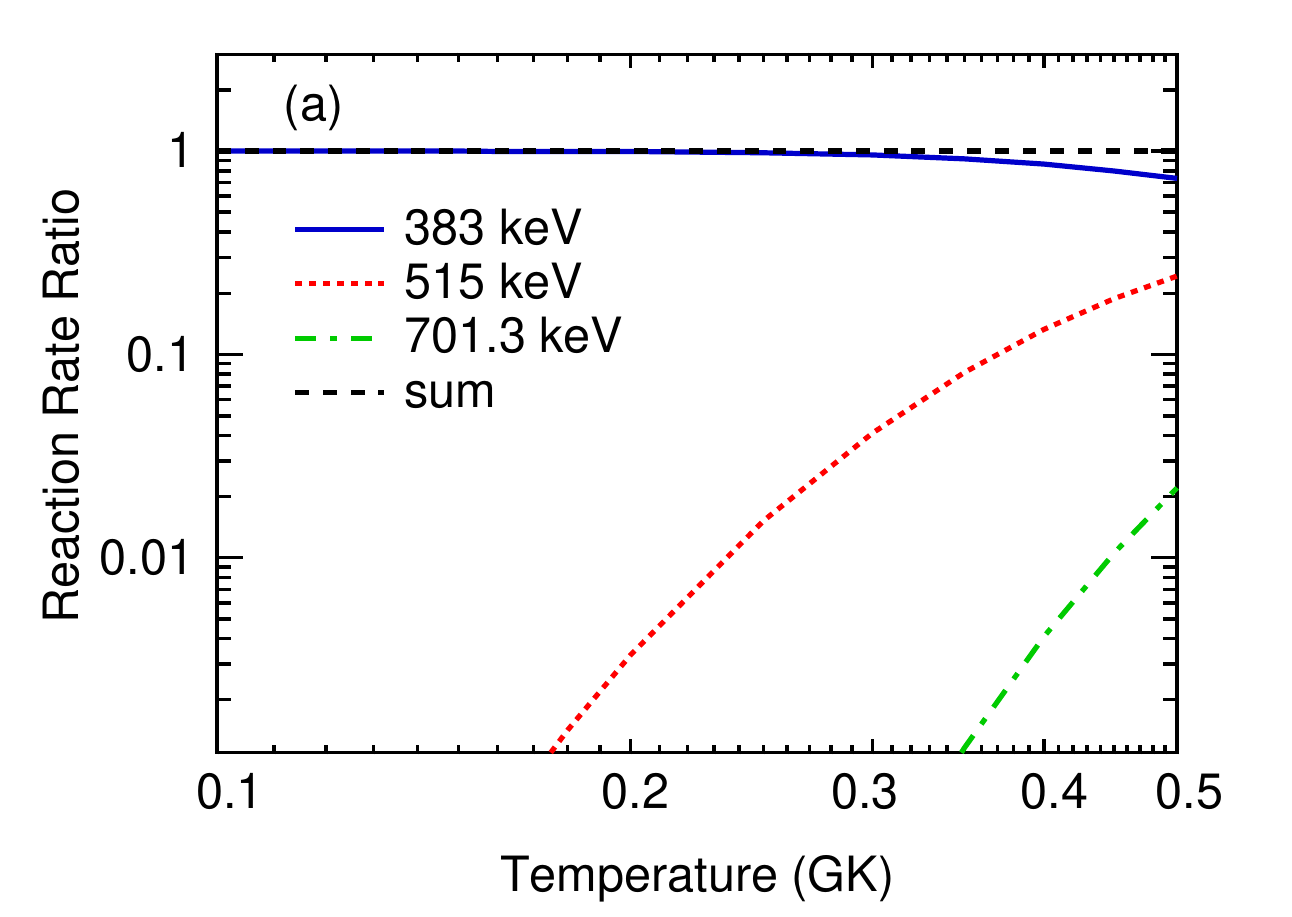}
    \label{figure7a}}
   \subfloat[Ratio of the new over old reaction rate.]{%
   \includegraphics[width=0.5\textwidth]{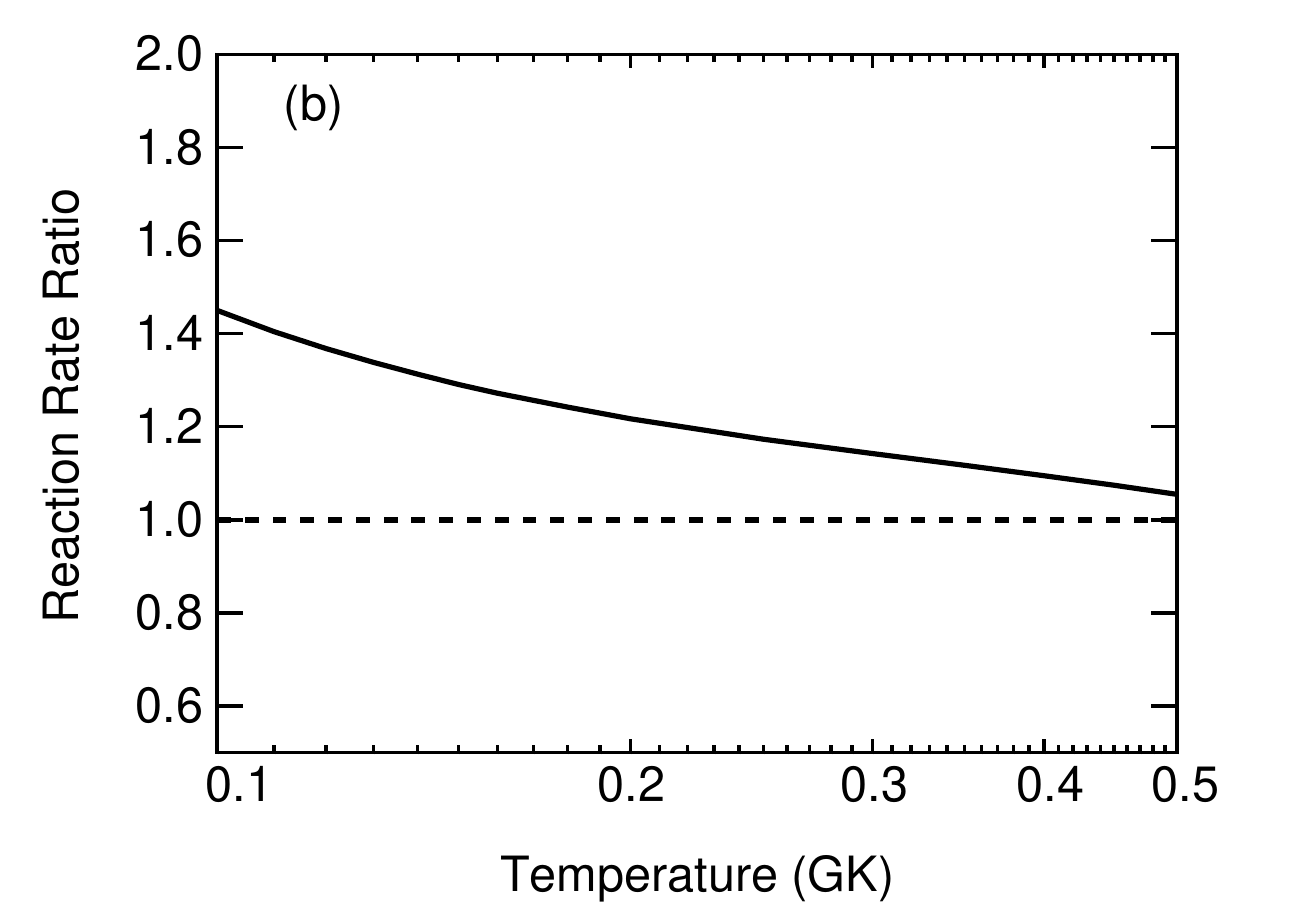}
    \label{figure7b}}
  \end{center}\vspace{-0.2cm}
\caption{\label{figure7}(Color online) (Left panel): The thick solid blue, curved dotted red and curved dash-dotted green lines are the ratios of upper limit resonance contributions for the 383-keV, 515-keV and 701.3-keV resonances, respectively, (see Table~\ref{tab:3}) to the sum of all three contributions. The latter (denoted by the straight dashed black line) is the total $^{38}$K($p, \gamma$)$^{39}$Ca resonant reaction rate, which is entirely dominated by the 383-keV resonance over the temperatures characteristic of explosive hydrogen burning in novae. Both axes are on logarithmic scale. (Right panel): The thick solid black line represents the ratio of the present total resonant $^{38}$K($p, \gamma$)$^{39}$Ca reaction rate to that of Ref.~\cite{Christian:2018}. The former is higher than the latter by up to a factor of 1.4. The x-axis is on logarithmic scale.}
\end{figure*}

\noindent ponding to the only resonance observed~\cite{Christian:2018}. We have thereby considered $\eta$ to be a constant over the small energy difference considered here. Finally, center-of-mass de Broglie wavelength can be calculated at each resonance energy.\par
The scaled resonance strengths for the 383-keV and 701.3-keV resonances are also given in Table~\ref{tab:3}. The uncertainties in these values come from propagating the 1$\sigma$ uncertainties on the $\omega\gamma$, $\epsilon$, and $E_{r}$ quantities from Ref.~\cite{Christian:2018} together with those on the present $E_{r}$'s and the scaled $\epsilon$'s. We have adopted the 515(10)-keV resonance from the excitation energy of 6286(10) keV measured in Ref.~\cite{Matoba:1993}, and its strength is adopted from the upper limit derived in Ref.~\cite{Christian:2018}. Lastly, for the $E_{r}$ $=$ 679$^{+2}_{-1}$(stat.) $\pm$ 1(sys.) keV and $\omega\gamma$ $=$ 120 $\pm$ 20(stat.) $\pm$ 15(sys.) meV values measured in Refs.~\cite{Lotay:2016,Christian:2018}, the statistical and systematic uncertainties are added together in quadrature to derive the final uncertainties in the resonance energy and strength for the 679-keV resonance.\par
For the 383-keV and 515-keV resonances in Table~\ref{tab:3}, we have assumed a spin/parity of 5/2$^{+}$ instead of 3/2$^{+}$ because the corresponding resonance strengths given in Ref.~\cite{Christian:2018} were obtained assuming the spin/parities of 5/2$^{+}$. Also, this assignment is consistent with an \textit{l} $=$ 0 transfer in the $^{38}$K $+$ $p$ system, whereas a $J^{\pi}$ $=$ 3/2$^{+}$ state in $^{39}$Ca corresponds to an \textit{l} $=$ 2 transfer, which makes the contribution of such a resonance to the $^{38}$K($p, \gamma$)$^{39}$Ca reaction rate less important.\par
With these, the total $^{38}$K($p, \gamma$)$^{39}$Ca resonant reaction rate was calculated analytically~\cite{Longland:2010}:
\begin{multline}
N_{A}<\sigma\upsilon>_{r}\,=\, \frac{1.5399\,\times\,10^{11}}{T_{9}^{3/2}}\Big(\frac{M_{0}\,+\,M_{1}}{M_{0}M_{1}}\Big)^{3/2}\,\times \\
\sum_{i}(\omega\gamma)_{i}\,\exp\Big(\frac{-11.605E_{i}}{T_{9}}\Big),
\label{equation4}
\end{multline}

\noindent where $N_{A}<\sigma\upsilon>_{r}$ is the resonant reaction rate (in cm$^{3}$\,mol$^{-1}$\,s$^{-1}$); $T_{9}$ is the temperature (in GK), $M_{0}$ and $M_{1}$ are the masses (in amu) of proton and $^{38}$K, respectively, ($\omega\gamma$)$_{i}$ is the strength of resonance $i$ (in MeV) and $E_{i}$ is the energy of resonance $i$ (in MeV).\par
Figure~\ref{figure7a} compares the ratio of individual upper limit resonance contributions for the resonances listed in Table~\ref{tab:3} (under the present work column) to the total recommended resonant rate. Under this assumption, the 383-keV resonance entirely dominates the $^{38}$K($p, \gamma$)$^{39}$Ca reaction rate over the nova temperature regime. While, the contributions of the 515-keV and 701.3-keV resonances to the total $^{38}$K($p, \gamma$)$^{39}$Ca resonant rate does not start until temperatures much higher than those of interest to novae.\par
The present total recommended $^{38}$K($p, \gamma$)$^{39}$Ca resonant reaction rate is compared with that of Ref.~\cite{Christian:2018} in Fig.~\ref{figure7b}. Even though the reaction rate is dominated by the 383-keV resonance over the nova temperature regime, and the energy of this resonance from the present work is only 3 keV lower than that used in Ref.~\cite{Christian:2018}, our recommended rate is higher in the temperature regime of interest for novae by at most a factor of 1.4 at 0.1 GK. This is because the present reaction rate is calculated based on scaled upper limit resonance strengths, and the contribution of the 515-keV resonance to the reaction rate starts to be significant (33\%) compared to that of the 383-keV resonance at 0.5 GK, which is beyond the temperature range of interest for the hottest novae. Since the 383-keV resonance energy is lower compared to that used for rate calculations in Ref.~\cite{Christian:2018}, the present reaction rate is higher than that of Ref.~\cite{Christian:2018} because the resonance energy enters equation (\ref{equation4}) with a negative sign.

\section{\label{Conclusions}Conclusions}

In the present work, we have presented the results of a charged-particle spectroscopy experiment to study the level structure of $^{39}$Ca using the $^{40}$Ca($^{3}$He, $\alpha$)$^{39}$Ca reaction measured with the Enge split-pole spectrograph at TUNL.\par
The level structure of $^{39}$Ca above the proton threshold at 5770.9 keV is important in determination of the $^{38}$K($p, \gamma$)$^{39}$Ca thermonuclear reaction rate at temperatures characteristic of explosive hydrogen burning in novae.\par
Our measured excitation energies of $^{39}$Ca states agree within 1 -- 2$\sigma$ with the results of previous measurements for most of the states. However, two of the states (at $E_{x}$ $=$ 5537-keV, and 6472.2-keV) do not agree with those previously measured.\par
The 5537-keV state is inconsistent with the 5588(10)-keV state measured in Ref.~\cite{Matoba:1993}. In that work, the state was only weakly populated and no excitation energy uncertainties were reported in the original work~\cite{Matoba:1993}. Its spin-parity assignment agrees well with that of the present 5537-keV state.\par
The excitation energy of the present 6472.2-keV state is consistent with the energies measured in Refs.~\cite{Matoba:1993,Doll:1976} but is 22 keV higher than that measured in Refs.~\cite{Lotay:2016,Christian:2018} assuming they are the same state.\par
In addition, we have observed a peak where the 6286-keV~\cite{Matoba:1993} state was expected to be populated with a 47-keV kinematic shift. $^{39}$Ca remains the most probable origin of this peak amongst the contaminants considered. However, we conservatively label it as unidentified in our analysis. Future measurements are warranted to confirm the origin of this peak.\par
We have also observed a tentative weak state at 5908 keV, which may be a new state in $^{39}$Ca. A recent charged-particle spectroscopy measurement on $^{39}$Ca~\cite{Liang:2018} with even higher energy resolution than that of the present work may shed light on these issues.\par
Spin-parities of a few of the proton resonances of $^{39}$Ca significant for nova nucleosynthesis have been determined in the present work from DWBA calculations using FRESCO~\cite{Thompson:1988}. The results agree for most but not all cases with the values previously determined in other measurements~\cite{Chen:2018}.\par
Finally, the total resonant $^{38}$K($p, \gamma$)$^{39}$Ca reaction rate was determined in the present work at temperatures characteristic of explosive hydrogen burning in novae using three resonances at 383-, 515- and 701.3-keV. The resultant rate is higher than that previously determined~\cite{Christian:2018} by up to a factor of 1.4 due to the fact that our resonance energy for the dominant resonance is lower than that used in Ref.~\cite{Christian:2018}. The strengths of the 383-, and 515-keV resonances are not measured yet. The present result for the $^{38}$K($p, \gamma$)$^{39}$Ca reaction rate intrinsically depends on the upper limit resonance strengths estimated in Ref.~\cite{Christian:2018}.\par
Regarding the present 701.3-keV resonance, the measurement of Ref.~\cite{Christian:2018} would have not been optimized for observing this resonance since it would have been located at the beginning of the DRAGON gas target, where the gas density is not uniform. In addition, the DRAGON's acceptance would have not been set to receive the potentially measurable  recoils from this resonance due to its distance from the center of the gas target.\par
Finally, we would like to highlight the fact that the other low energy resonances corresponding to the tentative 5908-keV, the 6001-keV and 6083-keV states could also be very important for determination of the $^{38}$K($p, \gamma$)$^{39}$Ca reaction rate at the nova temperature regime. They have been left out of the present rate calculation due to their unknown properties. Therefore, further high resolution study is merited to measure the spin-parities, proton and gamma widths of these proton resonances, and to look for potentially unobserved low spin resonances in $^{39}$Ca. We furthermore suggest that an independent nova model calculation to study the effects of the $^{38}$K($p, \gamma$)$^{39}$Ca reaction rate on nova abundances be postponed until these issues are resolved.

\section*{\label{Acknowledgment}ACKNOWLEDGEMENTS}

The authors would like to thank TUNL technical staff for their contributions. This material is based upon work supported by the U.~S.~Department of Energy, Office of Science, Office of Nuclear Physics, under Award Number DE-SC0017799 and under Contract No.~DE-FG02-97ER41041; as well as the Natural Sciences and Engineering Research Council of Canada.

%%%%%%%%%%%%%%%%%%
%% Bibliography %%
%%%%%%%%%%%%%%%%%%

\bibliographystyle{apsrev}
\bibliography{References}
\end{document}